\documentclass[conference]{IEEEtran}
\UseRawInputEncoding
\IEEEoverridecommandlockouts
\usepackage{algorithm}
\usepackage{cite}
\usepackage[normalem]{ulem}
\usepackage{amsmath,amssymb,amsfonts}
\usepackage{graphicx}
\usepackage{float}
\usepackage{textcomp}
\usepackage{algpseudocode}
\usepackage{xcolor}
\usepackage{subfig}
\usepackage{tabularray}
\def\BibTeX{{\rm B\kern-.05em{\sc i\kern-.025em b}\kern-.08em
    T\kern-.1667em\lower.7ex\hbox{E}\kern-.125emX}}
    
    \usepackage[letterpaper, left=0.75in, right=0.75in, bottom=1in, top=0.76in]{geometry}
\begin{document}

\title{A Novel Splitter Design for RSMA Networks\\}
% Improving RSMA Network Reliability Through an  Innovative Splitting Approach
% Boosting RSMA Network Reliability: A Novel Splitting Approach
% A New Splitting Technique to Increase RSMA Network Reliability 
% A Novel Splitting Approach to Boosting RSMA Reliability 
% A Novel Splitter Design to Boost RSMA Reliability

\author{
\IEEEauthorblockN{Sawaira Rafaqat Ali\IEEEauthorrefmark{1}, Shaima Abidrabbu\IEEEauthorrefmark{1}\IEEEauthorrefmark{2}, H. M. Furqan\IEEEauthorrefmark{2}, and H\"{u}seyin Arslan\IEEEauthorrefmark{1}}\\
\IEEEauthorblockA{\IEEEauthorrefmark{1}Department of Electrical and Electronics Engineering, Istanbul Medipol University, Istanbul, 34810 Turkey.\\
}
\IEEEauthorblockA{\IEEEauthorrefmark{2}Department of IPR and License Agreements, Vestel Electronics,  45030 Manisa, Turkey.\\
Email: sawaira.ali@std.medipol.edu.tr, shaima.abidrabbu@std.medipol.edu.tr, \\ Haji.Madni@vestel.com.tr, huseyinarslan@medipol.edu.tr}

}

\maketitle

\begin{abstract}
Rate splitting multiple access (RSMA) has firmly established itself as a powerful methodology for multiple access, interference management, and multi-user strategy for next-generation communication systems. In this paper, we propose a novel channel-dependent splitter design for multi-carrier RSMA systems, aimed at improving reliability performance. Specifically, the proposed splitter leverages channel state information and the inherent structure of RSMA to intelligently replicate segments of the private stream data that are likely to encounter deep-faded subchannels into the common stream. Thus, the reliability is enhanced within the same transmission slot, minimizing the need for frequent retransmissions and thereby reducing latency. To assess the effectiveness of our approach, we conduct comprehensive evaluations using key performance metrics, including achievable sum rate, average packet delay, and bit error rate (BER), under both perfect and imperfect channel estimation scenarios. 
 
\end{abstract}

\section{Introduction}
The rapid growth of mission-critical Internet of things (IoT) applications, such as industrial automation, autonomous vehicles, and smart cities, presents significant challenges for modern wireless systems. These applications require high reliability, low latency, and consistent availability, which conventional wireless technologies struggle to provide for ultra-reliable low-latency communication (uRLLC) \cite{8705373}. To address these demands, next-generation multiple access (NGMA) is essential for improving network capacity and supporting a large number of IoT devices \cite{9693417}.

Recently, rate-splitting multiple access (RSMA) has emerged as one of the candidates for NGMA, offering a promising solution to meet the high-capacity requirements. 
It outperforms traditional schemes such as orthogonal multiple access (OMA), non-orthogonal multiple access (NOMA), space division multiple access (SDMA), and physical layer multicasting by offering greater flexibility and effectively addresses challenges like imperfect channel state information at the transmitter (CSIT), user mobility, and diverse quality of service requirements \cite{9831440}.
RSMA networks, like all wireless systems, encounter a fundamental trade-off between latency and reliability, particularly in uRLLC and IoT applications. The literature addresses this challenge through two key approaches: short-packet transmission and retransmissions. In \cite{katwe2022rate} and \cite{xu2022rate2}, RSMA-based short packet communication is proven to outperform NOMA and SDMA systems. Likewise, in\cite{xu2022rate}, RSMA is introduced for short packet uplink communication, demonstrating its superiority over NOMA with respect to both throughput and error probability performance. RSMA is also shown as a potential approach for low-latency communication, particularly tailored for short-packet downlink communication, as explored in \cite{dizdar2021rate}.

\begin{figure*}[h]
      \centering
        \includegraphics[scale=0.56]{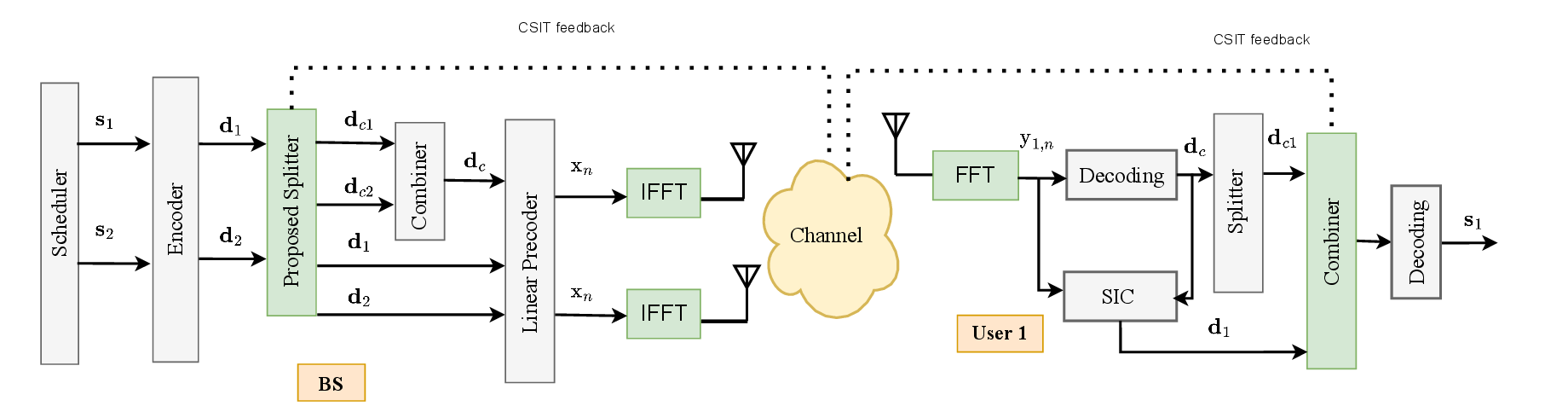}
        \caption{The proposed RSMA system model.}
        \label{fig1}
\end{figure*} 
From the retransmissions perspective, several studies have explored the integration of hybrid automatic repeat request (HARQ) protocols in RSMA networks. Abidrabbu et al. \cite{10315133} evaluated three retransmission strategies within the HARQ cycle. Loli et al. \cite{10606165} introduced a layered HARQ design for downlink RSMA, leveraging the common stream to manage retransmissions by sending new data alongside old data. Liu et al. \cite{10557543} developed an RSMA HARQ uplink protocol that optimizes retransmissions by avoiding the need to resend all failed streams. Additionally, Abidrabbu et al. \cite{10540604} proposed an incremental redundancy HARQ for RSMA networks, embedding private data redundancy into the common stream. Although RSMA enhances reliability through short-packet communication and HARQ, these methods have limitations. Specifically, short packets are suitable for low data rate IoT applications but are insufficient for high-throughput scenarios, and HARQ improves reliability at the cost of increased~delay.

From another perspective, traditional RSMA research has predominantly employed the concatenated form of data splitting, where data streams are divided using optimization techniques to meet specific requirements. For instance, the study in \cite{9967957} examines the secrecy performance of RSMA networks through optimal message division to enhance both sum and secrecy rates. In \cite{16}, the authors introduce an optimization framework for distributed rate-splitting method to maximize user data rates, while rate allocation and power control are addressed in \cite{9461768} to optimize sum rate under rate and successive interference cancellation (SIC) restrictions in multi-antenna systems. 

Unlike previous efforts, this study provides an intelligent splitter design for multi-carrier RSMA networks that improves reliability and reduces latency by utilizing the inherent structure of RSMA rather than typical optimization processes, short packet communication, or the HARQ protocol.  The contributions of this work are summarized as follows:

\begin{itemize} 
\item A channel-based splitter design is introduced for RSMA networks, aimed at improving reliability by utilizing CSIT and the intrinsic properties of RSMA, such as the higher power transmission for the common stream. Unlike traditional RSMA splitter, which divides the data into common and private streams, the proposed splitter intelligently duplicates portions of the private stream's data expected to encounter deep faded sub-channels and integrate them into the common stream. 
\item For the arrangement of selected private stream's data in the common, two methods are employed: the ``localized approach" and the ``distributed approach". In the case of the localized approach, the data corresponding to bad subcarriers for each user is placed in the common stream in concatenated form. Conversely, the distributed approach involves placing chunks of faded subcarriers' data for a user in a dispersed manner. 
\item To gain insight, key metrics such as bit error rate (BER), achievable sum rate and average packet delay are analyzed through a comparative study between the proposed RSMA approach and conventional techniques taking into account varying conditions, including perfect and imperfect CSIT, using precoders such as regularized channel inversion (RCI), and dirty paper coding (DPC).
\end{itemize}

\section{System Model And Proposed Approach}
Consider a single cell downlink multiuser RSMA network consisting of a base station (BS) and $K$ users indexed by the set $k=\{1,......, K\}$. The BS is equipped with $N_t \geq K$ transmit antennas, while each user is equipped with a single antenna. The communication channels are modeled as multi-path channels with $L$ exponentially decaying taps.
The data stream $\textbf{s}_k$ of the $k$-th user is encoded and mapped to $\textbf{d}_k$ which contains complex frequency domain data symbols given as:
\begin{equation}
\textbf{d}_k= \{d_{k,1},....,d_{k,N}\}, 
\end{equation} where $d_{k,n}$ presents a symbol of the $k$-th user on the $n$-th subcarrier and $N$ represents the total number of subcarriers. The $\textbf{d}_k$ stream is then passed through the proposed channel-dependent splitter as shown in Fig. \ref{fig1}\footnote{{
The associated multi-antenna system model in Fig. \ref{fig1} can simply be expanded to include more users and antennas.}}.
The proposed splitter takes the data stream, $\textbf{d}_k$, of $k$-th user and outputs a private stream and a common stream portion for $k$-th user. The data of the private stream is exactly the same as the input data which is $\textbf{d}_k$,
%The private stream is similar to input which is $\textbf{d}_k$,
while the user-specific portion of the common stream, $\textbf{d}_{ck}$,  contains a replica of those $m$ out of $N$ data symbols of  $\textbf{d}_k$ which are likely to encounter deep fading channels.
The selection of indices of these $m$ symbols out of $N$ symbols of  $\textbf{d}_k$ corresponding to deep-faded sub-channel is done based on the signal-to-noise ratio (SNR) for $k$-th user at $n$-th subcarrier which can be given as follows: 
\begin{equation}
\gamma_{k,n}=\frac{P|{H}_{k,n}|^2}{\sigma^2},
\end{equation} 
where $P$ is the power allocated to each subcarrier, $|{H}_{k,n}|$ is the magnitude of the frequency domain sub-channel, and $\sigma^2$ is the variance of additive white Gaussian noise (AWGN). To optimize the selection of $m$ indices, an equivalent  optimization problem is written as follows:
\begin{equation}
\label{3}
    \left\{z_{k,1}^{\text {opt }}, .., z_{k,m}^{\text {opt }}\right\}=\arg \min _{\left\{z_{k,1}, .., z_{k,N}\right\} \in \mathcal{Z}_{k,N}} \gamma_{\left[z_{k,1},.., z_{{k},{N}}\right]},
\end{equation}
where $\mathcal{Z}_{k,N}$ denotes the set of indices of $N$ subcarriers of private stream and $ z_{k,1}^{\text {opt }}, .., z_{k,m}^{\text {opt }}$ are the selected indices of subcarriers of the private stream that corresponds to low SNR. The symbols corresponding to these indices are replicated in a common stream. We consider a scenario where uniform power allocation is applied across all subcarriers, which simplifies the optimization problem to selecting $m$ subcarrier indices that correspond to the lowest channel gains. Therefore, (\ref{3}) can be rewritten as follows:
\begin{equation}
    \left\{z_{k,1}^{\text {opt }}, .., z_{k,m}^{\text {opt }}\right\}=\arg \min _{\left\{z_{k,1}, .., z_{k,N}\right\} \in \mathcal{Z}_{k,N}} |{H}|_{{\left[z_{k,1}, .., z_{{k},{N}}\right]}}.
\end{equation}

As a result, the proposed splitter outputs a user-specific portion of the common stream, \(\textbf d_{ck}\), along with a private stream, \(\textbf{d}_{k}\), for each user. The user-specific common stream's portions, \(\textbf d_{ck}\), are then directed to a combiner, where their data is arranged and encoded into one stream $\textbf{d}_c= \{\textbf{d}_{c1},\textbf{d}_{c2},....\textbf{d}_{cK}\}$ where $\textbf{d}_{ck}= \textbf{d}_{k}(z_{k,1}^{\text {opt }}, .., z_{k,m}^{\text {opt }})$. Two methods are used to organize the replicated private data within the common stream: the localized and distributed. In the localized approach, the data corresponding to the selected indices for each user is structured in the common stream in a concatenated format described as follows: 
\begin{equation}
\textbf{d}_{c}^{\text{loc}} = \{\underbrace{d_{z_{1,1}}, d_{z_{1,2}}, \ldots, d_{z_{1,m}}}_{\text{User 1 ($\textbf{d}_{c1}$)}}, \ldots, \underbrace{d_{z_{K,1}}, d_{z_{K,2}}, \ldots, d_{z_{K,m}}}_{\text{User K ($\textbf{d}_{cK}$)}}\}.
\end{equation}
On the other hand, in the distributed approach, the data is arranged in an interleaved format, allowing any combination of two or more symbols, illustrated as follows: 
\begin{equation}
 \textbf{d}_{c}^{\text{dis}} =\{ \underbrace{d_{z_{1,1}}, d_{z_{1,2}}}_{\text{User 1}},  \underbrace{d_{z_{2,1}}, d_{z_{2,2}}}_{\text{User 2}}, \ldots, \underbrace{d_{z_{1,m-1}}, d_{z_{1,m}}}_{\text{User 1}}, \ldots\}. 
\end{equation}
The transmitted signal at $n$-th subcarrier can be written as:
\begin{equation}
\label{1}
x_{n}= \mathbf{p}_{c,n} d_{c,n}+\sum_{k=1}^{K}
\mathbf{p}_{k,n} {d}_{k,n}, 
\end{equation}
where $\mathbf{p}_{c,n}$ and
$\mathbf{p}_{k,n}$
are the linear
precoders applied to the common stream and the private
stream of $k$-th user at $n$-th subcarrier and $d_{c,n}$ is the symbol of common stream at $n$-th subcarrier. 
Then, the signal is converted into a time domain by inverse fast Fourier transform (IFFT) operation and transmitted over the channel after cyclic prefix (CP) insertion. 

At the receiver, upon reception, the $k$-th user discards CP and then performs the fast Fourier transform (FFT) process to transform the signal into the frequency domain. The received signal of the $k$-th user at $n$-th subcarrier is given as: 
\begin{equation}
{y}_{k,n}=\mathbf{h}_{k,n}^{H}{x}_{n}+ {n}_{k,n}, 
\end{equation} 
where $\mathbf{h}_{k,n}$
is the channel vector of $k$-th user at $n$-th subcarrier and
\({n}_{k,n}\) follows a complex Gaussian distribution \(\mathcal{CN}\left(0, \sigma^2 \mathbf{I}\right)\), representing AWGN. 
Afterward, it decodes the common stream $\textbf{d}_c$ by treating all the private streams as noise. Subsequently, it decodes its private message $\textbf{d}_k$ by subtracting the common stream from the overall received signal using the SIC process. Following this, the splitter separates the user-specific common stream data $\textbf{d}_{ck}$ from the entire common stream and directs it to the combiner. Based on available CSIT, the combiner combines this data from the common stream which was replicated from the private stream with the corresponding data of subcarriers of the private stream that experienced deep faded sub-channels using maximum ratio combining. After this combining process, the original message $\textbf s_k$ is obtained. It's noteworthy that every user in the network follows the identical procedure.

In order to evaluate the performance of the proposed splitter approach in RSMA networks, the achievable sum rate improvement for all users is defined as follows:
\begin{equation}
\mathcal{W} [\mathrm{bps}/\mathrm{Hz}]={R_{c,n} +\sum_K R_{k,n}},
\end{equation}
where $R_{k,n}$ and $R_{c,n}$ is the rate for the private stream and common stream of the $k$-th at $n$-th subcarrier, respectively. $R_{k,n}$ can be calculated by taking the ensemble averages over $Z$ realizations of the channel, i.e.,
\begin{equation}
\begin{aligned}
{R}_{k,n} & =\frac{1}{Z} \sum_{z=1}^Z \log _2\left(1+\frac{|(\mathbf{h}_{k,n}^{(z)})^H \mathbf{p}_{k,n}^2|}{\sum^{K}_{i=1, i\neq k}|(\mathbf{h}_{i,n}^{(z)})^H \mathbf{p}_{i,n}^2|+\sigma^2}\right). 
\end{aligned} 
\end{equation}
To guarantee that all users can successfully decode the common stream, the rate of the common stream should be selected as in \cite{mao2018rate} given as
\begin{equation}
R_{c,n} = \min _{k\in K} \log_2 \left( 1 + \frac{| \mathbf{h}_{k,n}^H \mathbf{p}_{c,n}^2|}{\sum _{j=1}^{K} | \mathbf{h}_{k,n}^H \mathbf{p}_{j,n}^2| + \sigma_{k}^2} \right). 
\end{equation}

Notably, by taking advantage of the higher power transmission of the common stream, there is a high possibility to receive the data corresponding to the low channel gain successfully. Consequently, this approach introduces a diversity gain, significantly improving reliability within the same time slot, and minimizing the need for frequent retransmissions.

The computational complexity of the proposed method, analyzed under perfect CSIT for distributed and localized approaches (Table \ref{Table3}), increases with the number of users and subcarriers. Under perfect CSIT with the distributed approach, complexity rises by 48\% when users increase from 2 to 4 (64 subcarriers) and by 66\% when subcarriers double from 64 to 128 (2 users). The distributed approach exhibits higher complexity due to its dispersed data allocation, making it more demanding than the localized approach.
\begin{table}[H]
\centering
\caption{Computational Complexity Analysis.}\label{Table3}
\renewcommand{\arraystretch}{2}
\resizebox{1\columnwidth}{!}{
\begin{tblr}{
  cell{1}{2} = {c=4}{},
  cell{1}{6} = {c=4}{},
  cell{2}{2} = {c=2}{},
  cell{2}{4} = {c=2}{},
  cell{2}{6} = {c=2}{},
  cell{2}{8} = {c=2}{},
  hlines,
  vlines,
}       &  Distributed Approach &     &            &     & Localized Approach &     &            &     \\
\textbf{Number of users}                                         & \textbf{2}  &     & \textbf{4} &     & \textbf{2} &     & \textbf{4} &     \\
\textbf{Number of subcarriers}                                   & 64          & 128 & 64         & 128 & 64         & 128 & 64         & 128 \\
{\textbf{Perfect CSIT} } & 582         & 1195 & 959        & 1691 & 573        & 1116 & 629        & 1179 \\
% {\textbf{Imperfect CSI}} & 576         & 1180 & 987        & 1772 & 563        & 1103 & 626        & 1123 
\end{tblr}}
\end{table}

\begin{figure*}
\centering
\subfloat[]{\includegraphics[width=3.3in]{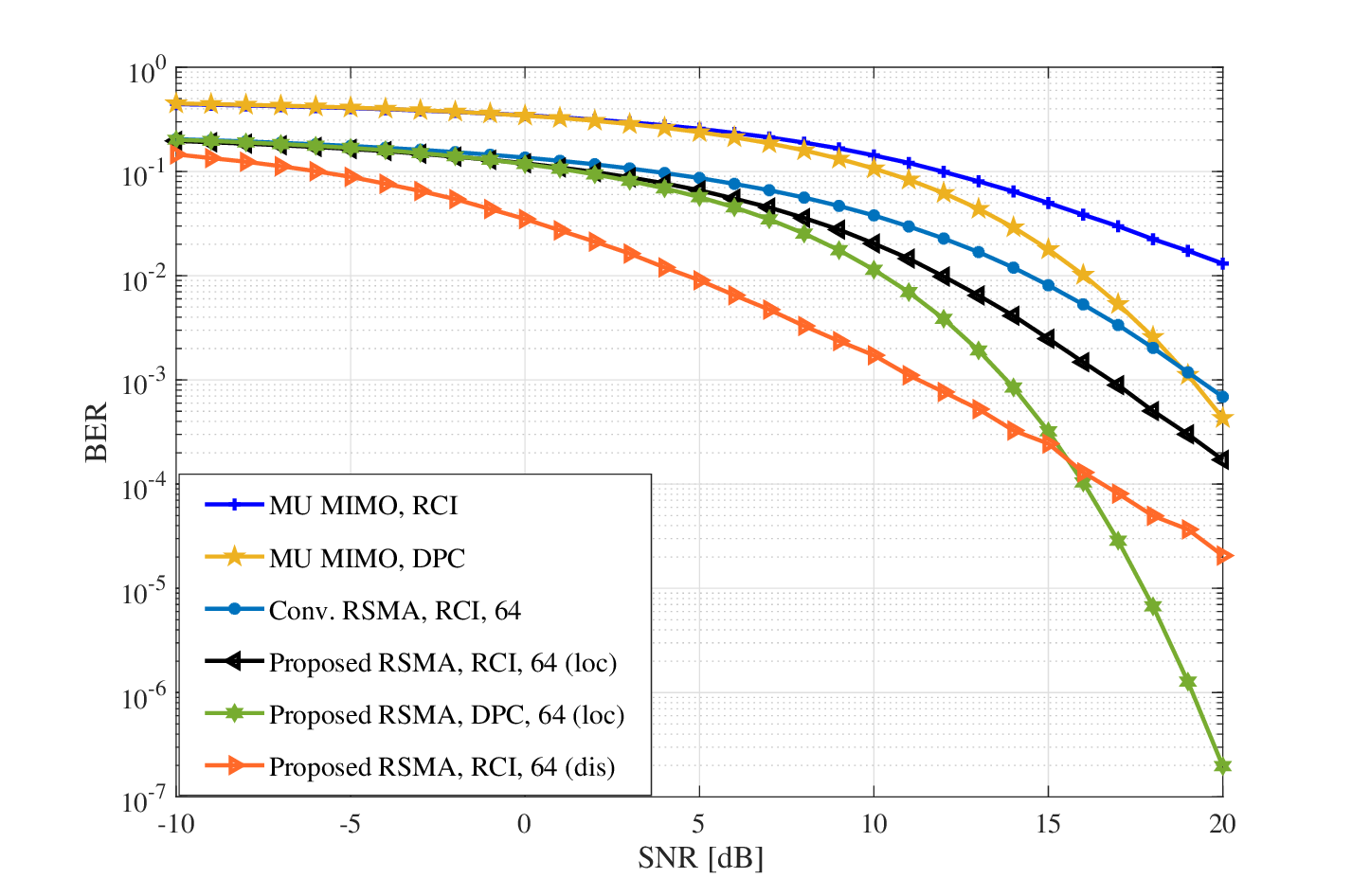}%
}
\hfil
\subfloat[]{\includegraphics[width=3.3in]{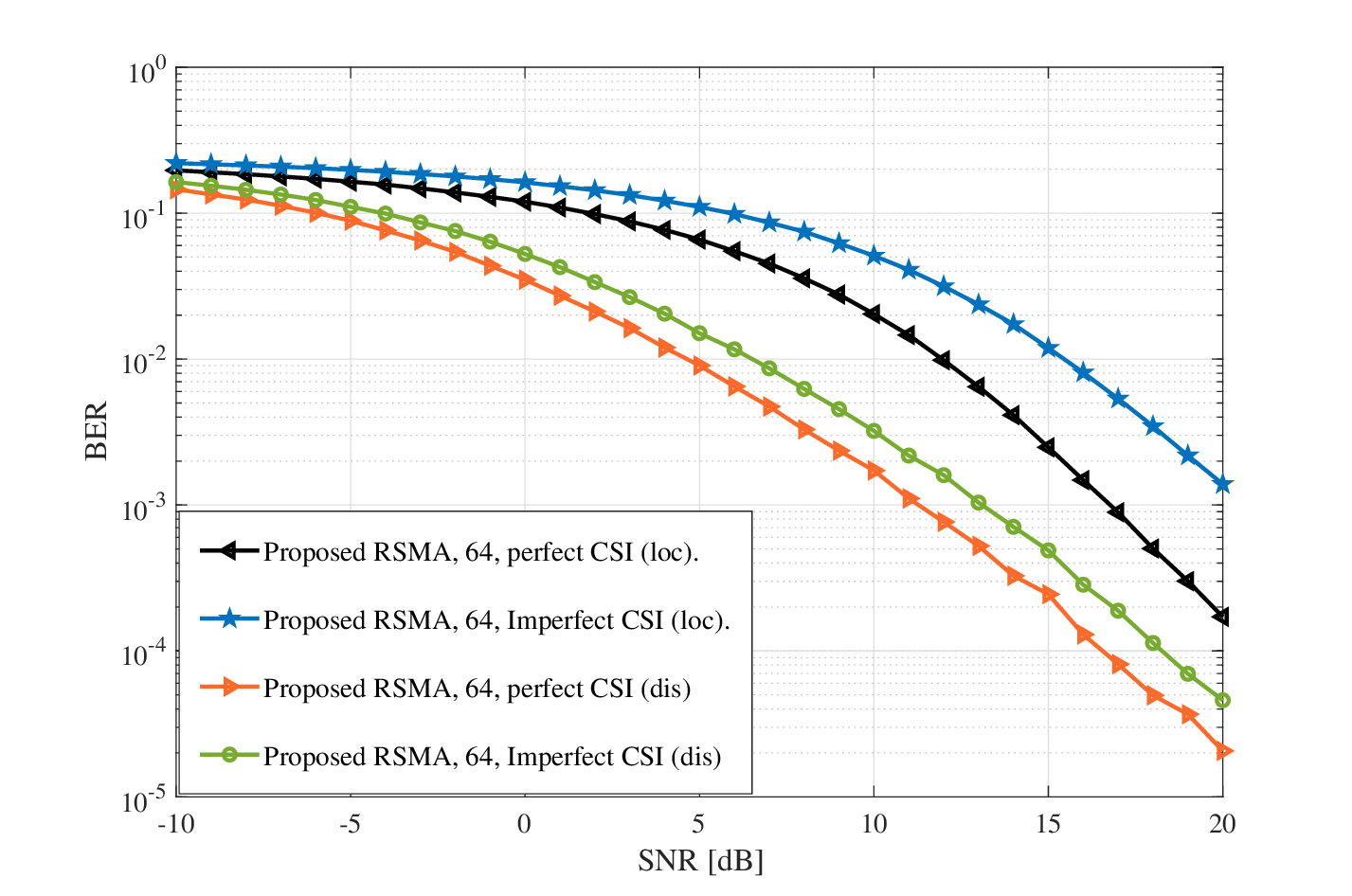}%
}
\caption{ BER evaluation: (a) of proposed RSMA under perfect CSIT, and (b) of proposed RSMA under imperfect CSIT.}
\label{fig331}
\end{figure*}

\section{Analytical and simulation results}
This section presents the analytical and simulation results of the proposed approach. Table \ref{Table2} outlines the simulation settings used to generate the results, while Algorithm 1 provides a detailed description of the overall proposed approach. 
\begin{table}[!h]
\caption{Simulation Parameters.}
\label{Table2}
\renewcommand{\arraystretch}{1.5}
\resizebox{1\columnwidth}{!}{
\begin{tabular}{|l|l|}
\hline
\textbf{Parameters} & \textbf{Assumptions} \\ \hline
Modulation type & BPSK \\ \hline
Noise density ($\sigma^{2}$) & {-80 dBmW/Hz} \\\hline 
%Number of users ($K$) &  2 \\ \hline 
Channel type &  {Rayleigh fading channel} \\ \hline 
Number of transmitter antennas ($N_t$) &  {2,4} \\ \hline 
% Number of receiver antennas ($N_r$) &  {1} \\ \hline  
Total number of sub-carriers ($N$) &  {64}, {128} \\ \hline 
The modulated block length ($S$) &  {20000} \\ \hline
Total number of channel realizations ($Z$) &  {1000} \\ \hline 
\end{tabular}}
\end{table}
\begin{algorithm}[H]
\begin{algorithmic}[1]
\caption{Overall Proposed Approach Algorithm.}
 \State \textbf{Input parameters}:
     $t=0$, $K$; $N$; $\mathbf{p}_{c,n}$; $\mathbf{p}_{k,n}$; $N_t$; and $S$. %number of active user, frames,
    % \State \textbf{Procedure}. 
    \For{$k=1$: $K$}
    \For{i=1: Length(SNR)}
        \For{$i_{packet}$=1:$S$}
        \State Generate (1)
        %\State {\textbf{Require:} CSIT availability, $K$, $N$, and $N/K$}
        \State {\textbf{Obtain} {(4) and (5) or (6)}} 
        %\State {\textbf{Obtain} {(5) or (6)}}
        \State {\textbf{Construct} $\textbf{d}_c$ and }{\textbf{Send it.}}
        %{(7)} over the channel $\mathbf{h}$}
        \State {RSMA receiver steps.}
        \EndFor  
\State BER calculations
\EndFor
  \State {\textbf{Update} {(4)}, e.g., $\textbf{h}(t) \neq \textbf{h}(t+1)$, and $N/K$}
\EndFor \\   
\textbf{Outputs:} BER and $\mathcal{W}$. 
\end{algorithmic}
\end{algorithm}

Fig. \ref{fig331}(a) shows the BER performance of the proposed RSMA considering localized and distributed approaches, together with conventional RSMA and multiuser MIMO (MU-MIMO) \cite{cho2010mimo} that employ different precoding strategies such as DPC and RCI under perfect CSIT. An evident trend in all techniques is the decrease in BER with increasing SNR. However, the proposed algorithm exhibits distinctive and superior behavior. In particular, at a given SNR, the BER reduction with different precoding approaches (DPC and RCI) is more pronounced for the proposed algorithm compared to conventional RSMA and MU-MIMO, which utilizes the same precoding techniques. This improvement can be attributed to the diversity gain achieved by replicating segments of private data encountering deep-faded channels within the common stream. Furthermore, the proposed strategy with a distributed approach under RCI precoding achieves the lowest BER among the other techniques due to its reduced susceptibility to burst errors.

To demonstrate the robustness of the proposed algorithm against imperfect CSIT, the impact of imperfections is considered, as depicted in Fig. \ref{fig331}(b). The figure demonstrates the robustness of the proposed approach in handling imperfect CSIT, attributed to the inherent advantages of RSMA. Particularly, RSMA proves to be a promising solution due to its resilience to CSIT inaccuracies. The reduced sensitivity of RSMA to CSIT imperfections further highlights its ability to operate effectively with partial CSIT. 

\begin{figure*}
\centering
\subfloat[]{\includegraphics[width=3.3in]{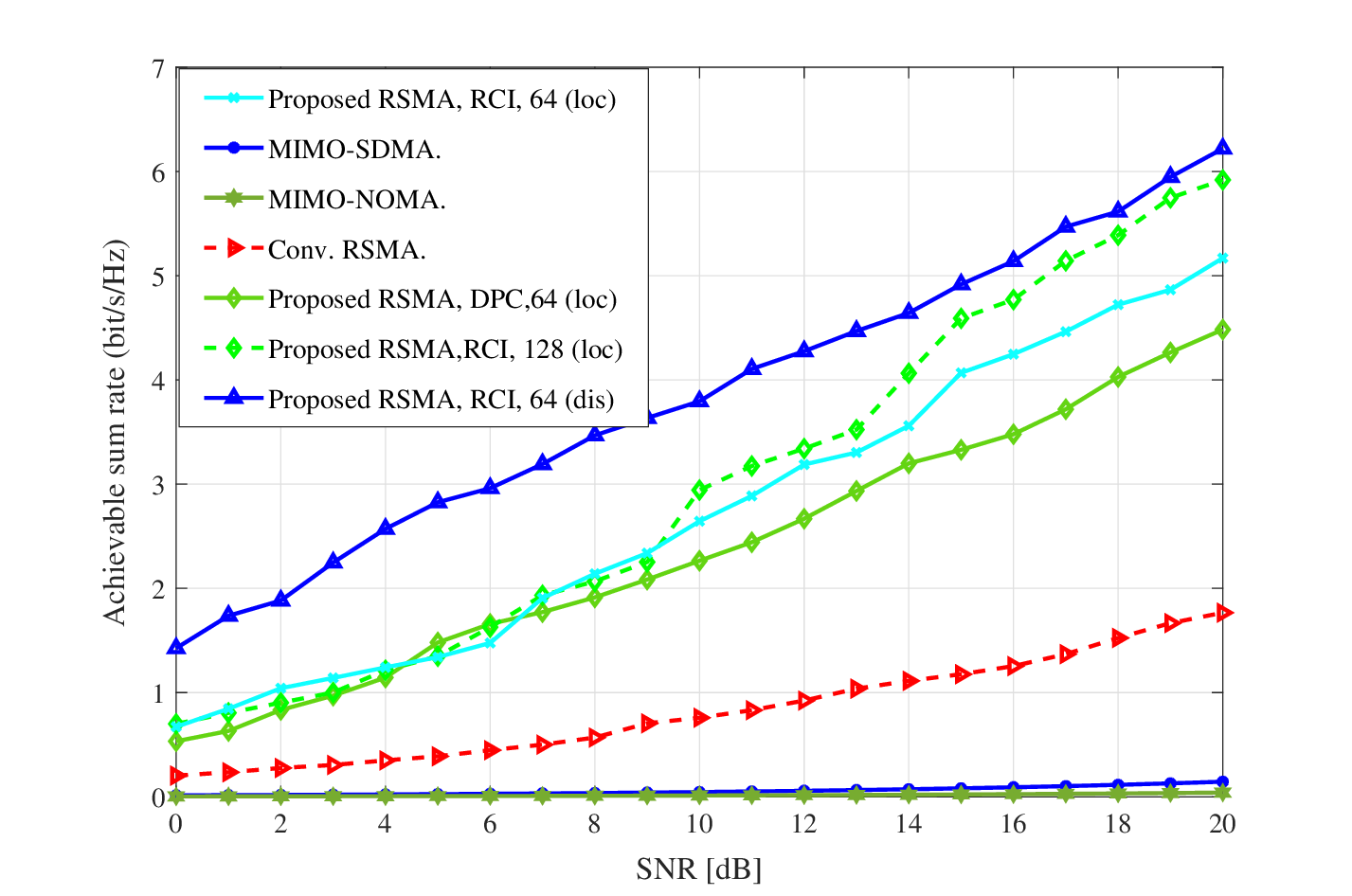}%
}
\hfil
\subfloat[]{\includegraphics[width=3.3in]{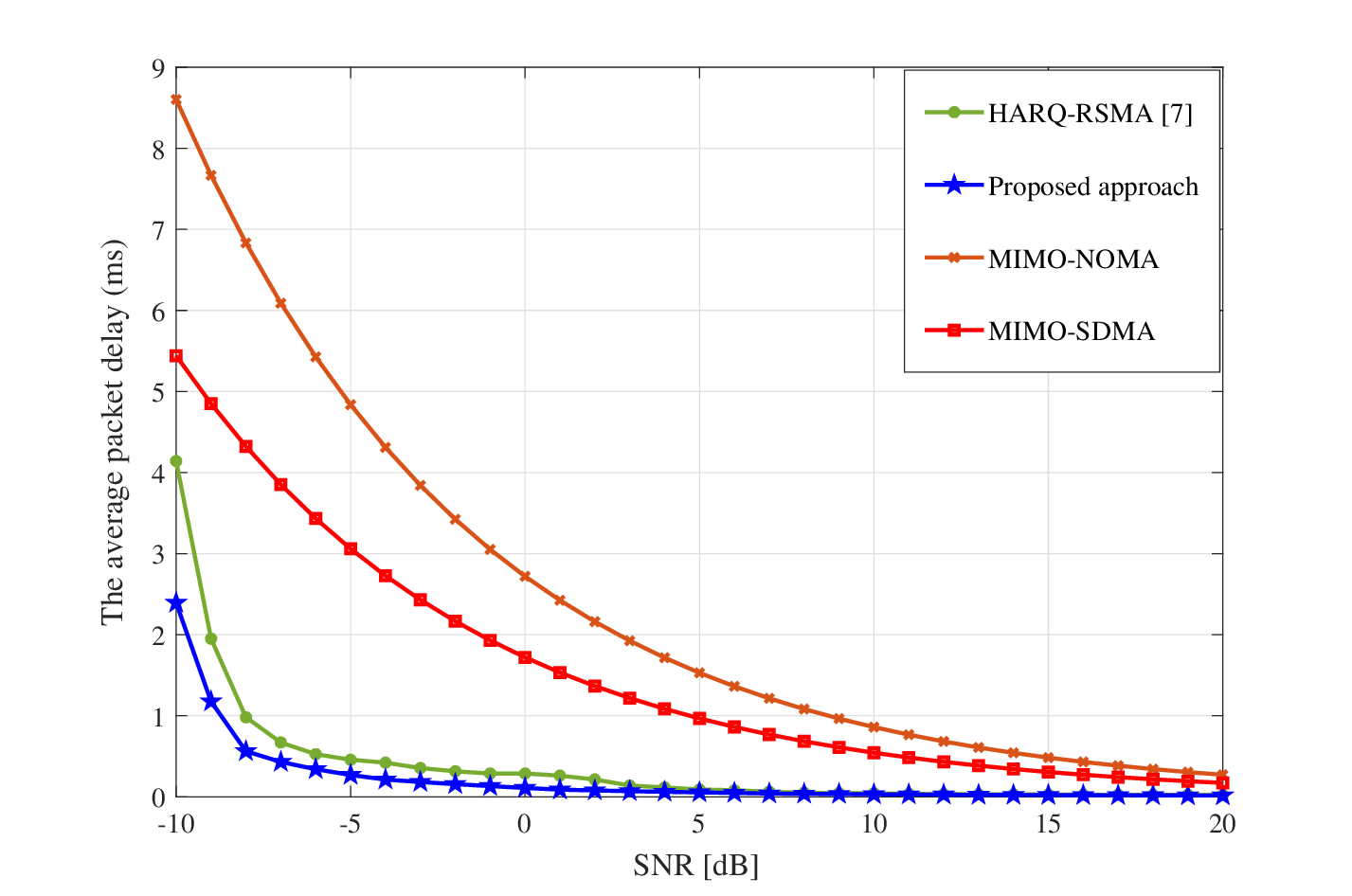}%
}
\caption{(a) The achievable sum rate of the proposed approach, and (b) the average packet delay of the proposed approach.}
\label{fig333}
\end{figure*} 
In Fig. \ref{fig333}(a), the performance of the achievable sum rate is depicted for the proposed RSMA scheme considering localized and distributed approaches, conventional RSMA, MIMO-NOMA, and MIMO-SDMA. The figure illustrates that the achievable sum rate increases with SNR. Specifically, the achievable sum-rate performance of the proposed RSMA approach under DPC and RCI with 64 subcarriers exceeds that of conventional RSMA, MIMO-NOMA, and MIMO-SDMA. This improvement can be attributed to the replication of a portion of the private data in the common stream, eliminating the need for retransmissions. As a result, the overall data transmission rate within a given time frame increases, leading to a higher achievable sum rate. Furthermore, it highlights that, when using 128 subcarriers, the achievable sum rate is further enhanced. In this case, users are provided with greater opportunities to transmit more diverse data via the common stream. In contrast, the distributed approach with RCI precoding demonstrates a higher data rate because of its reduced vulnerability to burst errors, which are more likely to occur in the localized approach. This immunity to errors contributes to the superior performance of the distributed method.

In Fig. \ref{fig333}(b), the average packet delay performance of the proposed approach is compared with that of conventional networks. The results demonstrate that the proposed approach achieves significantly lower average packet delay at -9 dB SNR, reducing delay by up to 40\% compared to the HARQ-RSMA method proposed in \cite{10315133}, 75.75\% compared to MIMO-SDMA, and 82.78\% compared to MIMO-NOMA. This improvement is attributed to the enhanced reliability achieved within the same transmission slot, which minimizes the need for frequent retransmissions and thereby reduces overall packet delay.   

\section{Conclusion and Future work}
This paper presents a new channel-dependent splitting approach for RSMA networks by leveraging CSIT and the inherent characteristics of the network. The proposed method improves transmission efficiency by allowing parts of the private stream, particularly those experiencing low channel gains, to be transmitted through the common stream. Utilizing the higher power feature of the common stream, the selected data can be accurately received within the same time slot, leading to increased reliability, reduced latency and a higher achievable sum rate. This approach is particularly beneficial in applications such as uRLLC, where reliability, low latency, and high sum rates are critical performance indicators.
Future research may investigate the adaptive subcarrier distribution strategies, such as adjusting the number of subcarriers to individual users and experimenting with different subcarrier arrangements inside a single stream to improve network performance.
%For future work, adaptive subcarrier allocation strategies could be investigated, tailoring the number of subcarriers to individual users and exploring diverse arrangements of subcarriers within the common stream to further optimize network performance.
\section*{Acknowledgement}
This work was supported by VESTEL and Technological Research Council of Turkey (TÜBİTAK) under Grant 5230073 with the cooperation of İstanbul Medipol University.
\bibliographystyle{IEEEtran}
\bibliography{IEEEabrv.bib,ref.bib}{}
\end{document}